# Modeling of Stick-Slip Behavior in Sheared Granular Fault Gouge Using the Combined Finite-Discrete Element Method

Ke Gao[1], Bryan J. Euser[1], Esteban Rougier[1], Robert A. Guyer[1,2], Zhou Lei[1], Earl E. Knight[1], Jan Carmeliet[3], and Paul A. Johnson[1]

[1]Geophysics, Los Alamos National Laboratory, New Mexico, USA

[2]Department of Physics, University of Nevada at Reno, Nevada, USA

[3]Department of Mechanical and Process Engineering, Swiss Federal Institute of Technology Zurich (ETH Zurich), Switzerland

Corresponding author: K. Gao (kegao@lanl.gov); P.A. Johnson (paj@lanl.gov)

**Key Points**

- The combined finite-discrete element method provides a powerful approach to modeling stick-slip behavior in granular fault gouge
- The number of slip events with large kinetic energy release and gouge layer thickness drop increases with increasing normal load
- As normal load increase, the earthquake reoccurrence (inter-event) time tends to be longer

**Abstract**

Sheared granular layers undergoing stick-slip behavior are broadly employed to study the physics and dynamics of earthquakes. Here, a two-dimensional implementation of the combined finite-discrete element method (FDEM), which merges the finite element method (FEM) and the discrete element method (DEM), is used to explicitly simulate a sheared granular fault system including both gouge and plate, and to investigate the influence of different normal loads on seismic moment, macroscopic friction coefficient, kinetic energy, gouge layer thickness, and recurrence time between slips. In the FDEM model, the deformation of plates and particles is simulated using the FEM formulation while particle-particle and particle-plate interactions are modeled using DEM-derived techniques. The simulated seismic moment distributions are generally consistent with those obtained from the laboratory experiments. In addition, the simulation results demonstrate that with increasing normal load, (i) the kinetic energy of the granular fault system increases; (ii) the gouge layer thickness shows a decreasing trend; and (iii) the macroscopic friction coefficient does not experience much change. Analyses of the slip events reveal that, as the normal load increases, more slip events with large kinetic energy release and longer recurrence time occur, and the magnitude of gouge layer thickness decrease also tends to be larger; while the macroscopic friction coefficient drop decreases. The simulations not only reveal the influence of normal loads on the dynamics of sheared granular fault gouge, but also demonstrate the capabilities of FDEM for studying stick-slip dynamic behavior of granular fault systems.

## 1. Introduction

Tectonic earthquakes generally occur due to a sudden release of the elastic energy accumulated in fault gouge, surrounding rocks and tectonic plates when subjected to long-time shear (Dorostkar et al., 2017b; Marone et al., 1990). The fault gouge, an ensemble of solid granular particles created by fragmentation and wearing of the fault blocks, plays a key role in the macroscopic sliding friction and the friction stability of the fault (Dorostkar et al., 2017b; Pica Ciamarra et al., 2011). Therefore, sheared granular layers undergoing stick-slip behavior are broadly employed to study the physics and dynamics of earthquakes and a number of laboratory experiments have been conducted in this regard (Annunziata et al., 2016; Geller et al., 2015; Johnson & Jia, 2005; Johnson et al., 2008; Marone, 1998; Marone et al., 1990; Passelègue et al., 2016a; Passelègue et al., 2016b; Scuderi et al., 2017a). For example, Geller et al. (2015) conducted an experiment using two photoelastic plates to compress and shear a granular gouge composed of nylon rods; Marone and colleagues devised an "earthquake machine", a bi-axial shear device, in which two layers of glass beads were sandwiched and sheared by three steel plates (Johnson et al., 2008; Marone, 1998; Scuderi et al., 2017a); Annunziata et al. (2016) developed an annular cell full of glass beads confined and sheared by an overhead rotating top plate.

Besides, with the rapid development of computer power and numerical approaches, various numerical modeling tools are employed to study the behavior of granular material (Dorostkar et al., 2017a, 2017b; Dratt & Katterfeld, 2017; Ferdowsi, 2014; Ferdowsi et al., 2013; Griffa et al., 2011; Griffa et al., 2012; Pica Ciamarra et al., 2011; Wang et al., 2017). In addition to the ease of implementation, numerical simulations of granular fault gouge allow for analysis of the mechanical behavior of the system at a level of spatial and temporal resolution not accessible experimentally (de Arcangelis et al., 2011), and also offer the possibility of identifying the most important parameters governing stick-slip dynamics (Dorostkar et al., 2017b). We note that all of these experiments essentially represent a single isolated fault patch, rather than a large number of them that would exist on a fault in the earth. Nonetheless, the physics of what is learned in experiments and simulations appears to have applicability to the real earth based on scaling relations such as Gutenberg-Richter (Gutenberg & Richter, 1955; Johnson et al., 2013) and applications by machine learning of laboratory approaches to the Earth, where fault physics can be inferred (Hulbert et al., 2018; Rouet-Leduc et al., 2018; Rouet-Leduc et al., 2017).

Among the many numerical methods used to describe the evolution of granular systems, the discrete element method (DEM), owing to its capability to consider the behavior of granular materials at the particle level, has demonstrated its value as a tool for such investigations and has broadened the understanding of stick-slip behavior in granular fault gouge (de Arcangelis et al., 2011; Dorostkar et al., 2017a, 2017b; Ferdowsi, 2014; Ferdowsi et al., 2013; Griffa et al., 2011; Griffa et al., 2012; Wang et al., 2017). In classic DEM models, granular fault gouge is usually represented by a pack of rigid particles, and the normal and tangential contact forces between particles are described using Hertzian contact and Coulomb friction, respectively (Dorostkar et al., 2017a); the representation of the shearing plate is often either ignored or simplified by a set of bonded particles (Abe & Mair, 2005; Dorostkar et al., 2017a, 2017b; Ferdowsi et al., 2013; Ferdowsi et al., 2014; Griffa et al., 2013; Mair & Abe, 2008; Wang et al., 2017) (Figure 1a). As a result, DEM is incapable of capturing real and detailed deformation and stress distributions within the particles and plates (Dratt & Katterfeld, 2017; Ma et al., 2016). Moreover, the deformation of these particles and plates provides the means for a wide spectrum of elastic energy storage and release, but for DEM the elastic energy can only be approximately interpreted based on particle overlap. Because the build-up and sudden release of elastic energy in particles and plates controls the stick-slip performance in granular media (Griffa et al., 2013), appropriate modeling of the real deformations of a granular system is critical to conduct an accurate exploration of the faulting process dynamics.

From a computational mechanics point of view, a granular fault system is essentially a combination of continua (each individual plate and particle) and discontinua (particle-particle and particle-plate interactions). Considering this, a numerical tool that has the capability of handling continua and discontinua simultaneously would be useful so that more can be learned regarding plate-gouge and particle-particle interactions. Fortunately, a recently developed numerical method – the combined finite-discrete element method (FDEM) (Munjiza, 1992; Munjiza, 2004; Munjiza et al., 2011; Munjiza et al., 2014), which merges finite element based analysis of continua with discrete element based transient dynamics, contact detection and contact interaction solutions of discontinua, provides a natural solution to modeling a fault system with hybrid characteristics. A simple FDEM realization of this type of fault system is presented in Figure 1b where each plate and each particle is represented by a discrete element, which allows for the tracking of their motion and their interactions with neighboring objects, and each discrete element is further discretized into finite elements in order to capture its deformation and stress evolution when subjected to external forces. A comparison between the DEM and FDEM model shown in Figure 1 demonstrates that the FDEM offers a more detailed and explicit representation of the granular fault system. To date, a systematic application of the FDEM to this kind of problem has not yet been found in the literature.

In this work, the FDEM is used to investigate the stick-slip behavior of a sheared fault containing gouge. Particular attention is given to the influence of different normal loads on the dynamics of slip events. In the following sections, we first provide the reader with a brief introduction to FDEM. Then, we illustrate the setup of the numerical model and provide a comparison of the simulated seismic moments with those obtained from the two-dimensional laboratory experiments conducted by Geller et al. (2015). Following this, we present how the friction coefficient, kinetic energy, fault gouge thickness and recurrence (inter-event) time between slips are influenced by different normal loads. The simulation capabilities of FDEM for granular fault system are explored, and we then present our conclusions.

## 2. The Combined Finite-Discrete Element Method (FDEM)

The FDEM was originally developed by Munjiza in the early 1990s to simulate the transition behavior of material from continuum to discontinuum (Munjiza, 1992). The essence of this method is to merge the algorithmic advantages of the DEM with those of the finite element method (FEM). The main theory of the FDEM can be mainly broken down into the following parts: governing equations, finite strain based formulation for deformation description, contact detection, and contact interaction algorithms (Lei et al., 2016; Munjiza & Andrews, 1998; Munjiza et al., 2006).

## 2.1. Governing Equations

The general governing equation of the FDEM is (Munjiza, 2004)

$$\mathbf{M}\ddot{\mathbf{x}} + \mathbf{C}\dot{\mathbf{x}} = \mathbf{f}, \qquad (1)$$

where $\mathbf{M}$ is the lumped mass matrix, $\mathbf{C}$ is the damping matrix, $\mathbf{x}$ is the displacement vector, and $\mathbf{f}$ is the equivalent force vector acting on each node which includes all forces existing in the system such as the forces due to material deformation and contact forces between solid elements. An explicit time integration scheme based on a central difference method is employed to solve Eq. (1) with respect to time to obtain the transient evolution of the system.

## 2.2. Finite Strain Based Formulation

In FDEM, each discrete element consists of a subset of finite elements that are allowed to deform according to the applied load. Deformation of the finite elements is described by a multiplicative decomposition based formulation (Munjiza et al., 2014). This framework allows for a uniform solution for both isotropic and general anisotropic materials for capturing detailed material deformation (Lei et al., 2016). Moreover, volumetric locking due to the lower order finite element implementations can be eliminated by using a selective stress integration scheme.

## 2.3. Contact Detection and Contact Interaction

The contact detection between discrete elements is conducted using the MRCK (Munjiza-Rougier-Carney-Knight) algorithm (Munjiza et al., 2011; Rougier & Munjiza, 2010). The MRCK contact detection algorithm is efficient which results in a theoretical CPU time proportional to the total number of contact couples, and it is applicable to systems consisting of many bodies of different shapes and sizes (Rougier & Munjiza, 2010).

Contact interaction is an important component of granular matter since it controls the constitutive behavior of the granular system. When contact couples are identified, a penalty function based contact interaction algorithm is used to calculate the contact forces between contacting elements (Munjiza, 2004; Munjiza et al., 2011). In the penalty function method, a small penetration or overlap is allowed between objects in contact, which then determines the normal contact force (magnitude and direction) acting on the contacting elements. In the present work, a "triangle to point" (Munjiza et al., 2011) contact interaction algorithm is used in which the target triangular element is discretized into a series of points distributed on its edges and each target point is considered as a Gauss integration point through which the distributed contact forces are integrated (Figure 2).

In terms of the normal contact force calculation, actual contact will not occur unless the target point is located inside of the contactor triangle. The normal contact force is calculated using the following equation obtained through a derivation in which the energy balance is preserved (Munjiza, 2004; Munjiza et al., 2011):

$$f_N \mathbf{n}_N = A E_p \frac{h}{H} \mathbf{n}_N, \qquad (2)$$

where $A = l_t / n_t$, $l_t$ is the length of the target element edge on which the target point is located, $n_t$ is the number of target points per element edge, $E_p$ is the penalty parameter, $h$ is the distance between target point and the contactor element edge, $H$ is the height of the contactor element associated with the contacting edge, and $\mathbf{n}_N$ is unit vector of the normal contact force (Figure 2).

The tangential contact force is assessed by means of the friction coefficient and the relative tangential displacement between a target point and contactor element occurring during a time step therefore taking into account the history of contact (Guises et al., 2009). The total relative tangential displacement between a target point and a contactor element accumulated at time step $t$, $d_t$, is recorded and used to calculate $d_{t+1}$:

$$d_{t+1}\mathbf{n}_T = \left(d_t + v_r^t \Delta t\right)\mathbf{n}_T, \tag{3}$$

where $\mathbf{n}_T$ is the unit vector of tangential force, $v_r^t$ is the relative tangential velocity between the target point and contactor element at time step $t$ and $\Delta t$ is the time step. The total tangential force at time step $t$ is calculated by

$$f_T^t \mathbf{n}_T = \left[\min\left(\mu f_N, AE_p \frac{d_t}{l_c}\right)\right]\mathbf{n}_T, \tag{4}$$

where $l_c$ is the length of the contactor element edge associated with tangential contact and $\mu$ is the frictional coefficient. The calculated contact forces are distributed among the nodes of contactor element and, in a similar manner, the opposite forces are applied to the target point and distributed among the nodes of the target element.

It is beyond the scope of the present paper to provide a detailed description of the above principles; however, details of these can be found in FDEM books (Munjiza, 2004; Munjiza et al., 2011; Munjiza et al., 2014). FDEM allows explicit geometric and mechanical realization of systems involving both continua and discontinua, which makes it superior to both pure FEM and DEM. Since its inception (Munjiza, 1992), FDEM has proven its computational efficiency and reliability and has been extensively used in a wide range of endeavors in both industry and academia, such as stress heterogeneity (Gao, 2017; Gao & Lei, 2018; Lei & Gao, 2018), permeability (Latham et al., 2013b; Lei et al., 2015), acoustic emission (Lisjak et al., 2013) and hydraulic fracturing (Viswanathan et al., 2015) in rock masses, tunneling (Lei et al., 2017; Lisjak et al., 2014), block caving (Lei et al., 2013; Vyazmensky et al., 2010) and rock blasting (Trivino & Mohanty, 2015) in rock mechanics, red blood cell aggregation in medicine (Xu et al., 2013), masonry wall stability (Reccia et al., 2012), coastal protection (Latham et al., 2009) and shell structure fracturing (Munjiza et al., 2013a). Particularly, in terms of granular flow and compaction, much work has been done regarding the validation, verification and application of FDEM for particles with different shapes in both 2D and 3D (Anastasaki et al., 2015; Guises et al., 2009; Guo et al., 2015; Latham et al., 2013a; Latham & Munjiza, 2012; Latham et al., 2008a; Latham et al., 2008b; Latham & Munjiza, 2004; Ma et al., 2016; Munjiza et al., 2013b).

Additionally, benefitting from the recent implementation of a large-strain large-rotation formulation and grand scale parallelization in FDEM by the Los Alamos National Laboratory (Lei et al., 2014; Munjiza et al., 2014), the FDEM software package – HOSS (Hybrid Optimization Software Suite) (Knight et al., 2015; Munjiza et al., 2013b) – offers a powerful tool to study the behavior of granular fault gouge.

## 3. Numerical Model Setup

Figure 3 illustrates the geometry of the FDEM model, which is based on the two-dimensional, photoelastic shear laboratory experiment conducted by Geller et al. (2015). Here, for simplicity, two-dimensional plane stress conditions are assumed. The model consists of 2,817 non-destructive circular particles (arranged in 9 rows and 313 columns, before consolidation) confined between two identical plates. The diameter of the particles is either 1.2 or 1.6 mm, and there are the same number of particles for both sizes which are randomly generated and placed in the space between the two main plates. The number of particles is close to the laboratory experiments in which 3,000 particles are used (Geller et al., 2015). According to the existing literature on simulation of stick-slip problems (Dorostkar et al., 2017b; Ferdowsi, 2014), the number of particles adopted in this work is appropriate for such investigation.

Each main plate has dimensions of $570 \times 250$ mm in width and height, respectively. At the interface between the main plates and the particles, a series of half-circular shaped "teeth" are added to the plates to increase the friction between plates and particles and, thus, facilitate the transmission of shear stresses (Figure 3). The diameter of the teeth and the distance between adjacent teeth are equal to the diameter (1.6 mm) and radius (0.8 mm) of the larger particle, respectively. A total of 286 "sensor" points (two rows of $N_s = 143$) are set on the half-circular teeth centers to track the seismic moment as well as the change in gouge layer thickness. Two elastic "foam-like" particles are placed on either end of the gouge layer in order to hinder particles from escaping. To avoid significant distortion of the plates due to shearing, two stiff bars are attached to the bottom end of the bottom plate and to the top part of the top plate; the normal load $P$ and shear velocity $V$ are applied directly to these stiff bars.

The particles have a Young's modulus of 0.4 GPa and Poisson's ratio of 0.4, while the Young's modulus and Poisson's ratio of the main plates are 2.5 MPa and 0.49, respectively. These material parameters are similar to the laboratory experiments (Geller et al., 2015). The soft plates make it possible to visually capture the immediate response of the plates with respect to slip vibration and stress evolution (see supplementary animation). The particle-particle and particle-plate friction coefficients are set to 0.15 based on previous studies that showed that a smaller friction coefficient allows for larger slip events and reduces the frequency of small fluctuations in the macroscopic friction signal (Ferdowsi, 2014). In terms of the penalty parameter between contacts (e.g. particle-particle and particle-plate contact), it theoretically should be infinity in order to avoid penetration between contacting elements; however, a large penalty parameter will yield a significantly small time step. Recently study shows that in general a penalty parameter that is about 1-2 orders of the Young's modulus can ensure the computational correctness (Tatone & Grasselli, 2015). By compromising between achieving the correct elastic response between contact elements and maximizing the time step size in order to reduce the overall computational expense, a penalty parameter ten times larger than the particles' Young's modulus, i.e. 4 GPa, is used.

In the first phase of the simulation, the granular fault gouge is consolidated by moving the top and bottom stiff bars towards each other. The top stiff bar is then fixed in space, and the normal load acting on the bottom stiff bar is gradually increased until it reaches a target value $P$. The entire model then undergoes a dynamic relaxation phase where the kinetic energy of the system is gradually reduced while the system settles. The model is considered to have reached equilibrium when the kinetic energy has stabilized. After consolidation, the thickness and length of the granular fault gouge are around 11.7 mm and 500 mm, respectively. At this point, the top stiff bar begins shearing with a constant horizontal velocity $V$, while the normal load $P$ on the bottom stiff bar is maintained throughout the simulation. In order to alleviate distortion of the model and to assure the effectiveness of the normal load and shear motion on the granular fault gouge, during the shearing stage the top stiff bar is allowed to move only in the $x$-direction and the bottom stiff bar is allowed to move only in the $y$-direction.

Because of the soft plates, a series of relatively small normal loads ranging from 12 kPa to 44 kPa, in increments of 8 kPa, are employed to investigate the influence of different loading conditions on the stick-slip behavior of the gouge. The normal loads used here are also close to the laboratory experimental studies (Geller et al., 2015). Because the shear velocity plays a pivotal role in the slip events in sheared granular fault gouge (Anthony & Marone, 2005; Dorostkar et al., 2017b; Ferdowsi, 2014; Marone, 1998), a series of trial runs are conducted to identify the appropriate shear velocity to use in the virtual experiments. Based on these, a shear velocity of $V = 0.5$ mm/s is selected, which guarantees the occurrence of slip events. This relatively higher shear velocity is used to speed up the shearing by considering the computational expense as well as to capturing more slip events (laboratory experiments use a shear velocity of ~4 µm/s).

In the FDEM model, six-node composite triangular elements are employed for the plates in order to avoid volumetric locking caused by the high Poisson's ratio (0.49), while three-node constant-strain triangular elements are used for the particles and the foams. A total of 12,940 six-node and 67,800 three-node triangle elements are generated in the entire simulation domain. Each particle is composed of 24 approximately equal size elements. This allows for each particle to maintain an approximately circular shape after meshing and yields a particle area that is about 95.5% of the corresponding circle (see inset of Figure 3), which is sufficient to capture particle deformation while also assuring the model is not too computationally expensive. The finite elements used in the present study are assumed to follow an isotropic elastic relationship and thus the objects such as particles and plates behave as elastic materials. The simulations use a time step of 1.0E-7 s, and are run on a parallel cluster utilizing 208 processors. The inertial number of the system can be calculated using

$$I = \frac{\dot{\gamma}d}{\sqrt{P/\rho}} \tag{5}$$

to quantify the significance of dynamic effects in a granular material (MiDi, 2004). Here $\dot{\gamma}$ is the shear rate, i.e.

$$\dot{\gamma} = \frac{V}{H_\mathrm{m}}, \tag{6}$$

where $V$ is the shear velocity and for the soft system used here $H_\mathrm{m}$ is the height of the model which is approximately 520 mm, $d$ is the average particle diameter, $P$ is the pressure and $\rho$ is the density. The parameters of the fault system give an inertial number between 5.9E-8 and 1.1E-7, which guarantees that the simulation falls into the quasi-static flow regime. Detailed material and calculation parameters are tabulated in Table 1.

## 4. Simulation Results

### 4.1. Model Evolution and Seismic Moment Comparison with Laboratory Experiments

The model for each normal load scenario is run for roughly 3.0E+8 time steps and its computation time is around 300 hours on the cluster employed. The total shearing time in each simulation is approximately 30 s with the model reaching steady state after the first 3 s approximately. This shearing duration is comparable to other similar DEM simulations (Dorostkar et al., 2017a, 2017b) and gives a total shear displacement of roughly 20 mm (including the pre-steady state), which is close to the laboratory experiments. As the shearing proceeds, deformation accumulation in the granular fault system is captured by the model, as can be seen in Figure 4a the distribution of major principal stress at the final state for the model subjected to 44 kPa normal load. To facilitate comparison, we further sketch in Figure 4b an overlay of the model outlines for the beginning and final states according to their coordinates to demonstrate the model deformation. The model deformation is also reflected in the fault gouge. Particularly, at the time immediately before a slip event, a distinct stress concentration can be visualized between particles (Figure 5a). While after the slip event, because of the release of elastic energy, deformations in the fault gouge become relatively uniform, and as a result, less stress concentration can be found (Figure 5b).

Since mainly seismic moment analyses have been reported for the laboratory experiments (Geller et al., 2015), in order to compare the numerical simulations with the laboratory experiments, the sensor positions recorded along the boundary between the upper shearing plate and fault gouge have been used to obtained the seismic moments, and they are further analyzed using the same approaches as the ones used in Geller et al. (2015) to evaluate the capability of FDEM for such investigation. After the model reaches steady state, the positions of the sensors are recorded every 20,000 time steps (i.e. every $\Delta t = 2$ ms). This time step interval for output recording is carefully chosen through a series of comparisons by considering the resolution of output as well as to avoid unnecessary noise. As the plates are sheared, the sensor points move accordingly to follow the elastic response of the system. Seismic moment analyses of the experimental set up are reported by Geller et al. (2015), where the moment requires the differential motion of each sensor in the $x$-direction between adjacent time step intervals. The differential motion of sensor $i$ at time $j$ is defined as

$$s_i^j = \left(x_i^j - x_i^{j-1}\right) \cdot \theta\left(x_i^j - x_i^{j-1} - \delta\right), \tag{7}$$

where $\theta(x)$ is the Heaviside function such that

$$\theta(x) = \begin{cases} 0, & x < 0 \\ 1, & x \geq 0 \end{cases}, \tag{8}$$

and the noise threshold $\delta$ is chosen in a similar manner to the laboratory experiments, i.e. $\delta = 1.8 \cdot V \Delta t = 1.8\text{E}-3$ mm. Differential motion values below this threshold are set to zero, and the number of non-zero contributions is defined as

$$N^j = \sum_{i=1}^{N_s} \theta(s_i^j - \delta). \tag{9}$$

The spatially integrated moment at time $j$ is

$$M^j = \sum_{i=1}^{N_s} s_i^j. \tag{10}$$

Based on the moment calculation, events can be categorized as "coherent" (C) (i.e. large and system spanning events) or "non-coherent" (NC) (i.e. small and localized events) (Geller et al., 2015). The variation in event classification arises from the heterogeneous force distribution in the granular fault gouge set up by stress chains (e.g. Figure 5). To extract more detailed spatial information, the center for an event at time $j$ is defined by the displacement-weighted sensor point position

$$X^j = \left(\sum_{i=1}^{N_s} s_i^j x_i^0\right) \Big/ M^j, \tag{11}$$

where $x_i^0$ is the nominal initial position of the sensor $i$ and here the sensor positions at time 3 s (i.e. at the beginning of steady state) are used. The spatial extent of event $j$ is proportional to the radius of gyration, i.e.

$$R^j = \sqrt{\sum_{i=1}^{N_s} s_i^j \left(X^j - x_i^0\right)^2 \Big/ M^j}. \tag{12}$$

A normalized quantity that reflects the spatial coherence of an event is

$$C^j = R^j \Big/ \left(\sqrt{12} N^j\right). \tag{13}$$

Using the criterion in Geller et al. (2015) that events with $C^j < 2$ and $M^j/N^j > 0.003$ are considered as C events, we present both the C and NC events in Figure 6. The results are similar to the laboratory experiments. That is, for NC events, $M^j/N^j$ is nearly constant with respect to $M$, which means the events differ in moment owing to the number of sensor points $N^j$ involved rather than the distance that they move; for C events, $M^j/N^j$ is linear in $M$ because the number of sensor points participating in an event is limited by the size of overall sensor points, i.e. $N^j \approx N_s$, so that $M^j/N^j$ increases in proportion to the average sensor point displacement. It is also apparent from Figure 6 that as the normal load increases, there are more C events with larger moment $M$ generated.

Figure 7 illustrates the probability density distributions of all, NC and C events. The results agree with the experimental data collected by Geller et al. (2015). The probability density distribution of all events $p(M)$ is consistent with the Gutenberg-Richter law (Gutenberg & Richter, 1955) and is predicted to scale as $M^{-3/2}$ (the power -3/2 is within the range of -1.4 to -1.8 observed in real earthquake (Geller et al., 2015)) (Figure 7a). The NC events are generally distributed as a power law consistent with $M^{-3/2}$ (Figure 7b). The C events are broadly peaked, concentrated at large $M$ with a probability density distribution that consistent with log-normal (Figure 7c). As the normal load increases, the probability of NC events decreases, while the C events become more probable, increasing at the expense of NC events and having broader distributions.

The simulated results using FDEM are not only consistent with the laboratory experiments, but also agree with our observations of increased likelihood of large seismic moments at higher normal load (Geller et al., 2015). This also demonstrates that it is the capability of FDEM for explicit plate deformation simulation that makes the seismic moment analysis possible. In addition to seismic moment, the information of friction coefficient, kinetic energy, and granular fault gouge layer thickness change can also been extracted from the simulations. In the next section, the stick-slip behavior of the granular fault gouge in terms of these indicators are presented. Since changes in seismic moment are essentially reflected in the variation of kinetic energy, kinetic energy is mainly presented as a representation of the kinematic behavior of the granular fault system in the rest of the paper.

## 4.2. Statistics of Friction Coefficient, Kinetic Energy and Fault Gouge Layer Thickness

The shear and normal forces between the main plates and particles, and the kinetic energy of the entire model are also recorded every 20,000 time steps. The shear and normal forces are calculated by first resolving the normal and tangential contact force between each particle-plate contact pair into *x*- and *y*-directions and then integrating them separately along the particle-plate interface. In other words, the shear and normal forces between the main plates and the particles are respectively the summation of particle-plate contact forces in the *x*- and *y*-direction. The ratio of the shear to normal force is then calculated as the macroscopic friction coefficient between the plates and granular fault gouge.

Besides, in the simulation, the *x*- and *y*-velocity of each node are calculated at each time step, thus the kinetic energy of the entire system at a specific time step is calculated by

$$E_k = \sum_{i=1}^{n} \frac{m_i}{2} \left( v_{x_i}^2 + v_{y_i}^2 \right), \tag{14}$$

where $m_i$, $v_{x_i}$ and $v_{y_i}$ are respectively the nodal mass, and nodal velocity in *x*- and in *y*-direction for the node *i*, and *n* is the total number of nodes in the system. Since plane stress is used in the simulation, the model is assumed to have 1 mm thickness in *z*-direction (i.e. one unit under the current unit system used in the model) when calculating the kinetic energy. Unlike the previous work conducted using DEM that only calculated the kinetic energy of the granular gouge (Dorostkar et al., 2017a, 2017b; Ferdowsi et al., 2013; Ferdowsi et al., 2014), here we report the kinetic energy of the entire system because the plates and particles are working together as an entire entity and it is the energy evolution in this entity that governs the stick-slip behavior in granular fault gouge.

Additionally, the *y*-positions of the sensor points mentioned above are extracted to calculate fluctuations in the gouge layer thickness. The gouge layer thickness is calculated every 20,000 time steps by summing over the *y*-coordinates of the points located on the upper and bottom plates, respectively, and then averaging their differences. The macroscopic friction coefficient, kinetic energy, and gouge layer thickness are all used as indicators for identifying slip events. In the following, we present the time series of the macroscopic friction coefficient, kinetic energy and gouge layer thickness, and calculate their statistics.

Figure 8a-e shows the macroscopic friction coefficient, kinetic energy, and granular fault gouge layer thickness change with respect to time for normal loads between 12 and 44 kPa, in which similar patterns of stick-slip behavior in the granular fault gouge are observed. In the stick phase, the macroscopic friction coefficient increases in an approximately linear manner before approaching the peak. At the end of the stick phase, there is a rapid drop of the macroscopic friction coefficient accompanied by a sudden release of kinetic energy, signifying a slip event. As the shearing continues, stick-slips are periodically generated. The change in granular gouge layer thickness demonstrates the dilation of the gouge layer during the stick phase due to particle rolling, and compaction when slip occurs attributed to particle rearrangement. The small fluctuations observed in the stick phase are indicators of micro-slips which have been observed in laboratory experiments as precursors to stick-slip failure in sheared granular materials (e.g. Johnson et al., 2013).

In order to compare the effect of the normal load on the macroscopic friction coefficient, kinetic energy and gouge layer thickness, the statistical characteristics of these three stick-slip indicators are calculated and the results are presented in Figure 9 in the form of boxplot. A box represents the first (Q1) and third (Q3) quartiles and its length is the interquartile range (IQR, i.e. IQR = Q3 – Q1), with the line in the middle being the median (second quartile, Q2), and the lower and upper whiskers representing (Q1 – 1.5 × IQR) and (Q3 + 1.5 × IQR), respectively.

As can be seen in Figure 9a, the macroscopic friction coefficient hovers mainly around 0.35-0.50, and does not show much variation with respect to varying normal load. The kinetic energy of the granular fault system increases with the increasing normal load (Figure 9b), which could be caused by more external work having been done on the fault system when a higher normal load is applied. Thus, a large amount of external work is converted into kinetic energy and results in a more energetic system, which is consistent with the seismic moment results presented in Figure 7. The granular fault gouge is more compact under higher normal loads, and this is supported by the decreasing gouge layer thickness shown in Figure 9c. All these observations suggest that the increasing normal load imposes more constraints on the gouge and yields a denser and well-contacted gouge layer. In a more compact fault system, there are only small changes allowed in particle contact and rearrangement and, therefore, only small variations in both the kinetic energy and gouge layer thickness can be triggered by stick-slip failure, as shown in Figure 9b and c, respectively.

In the following, the kinetic energy release, decrease of macroscopic friction coefficient and gouge layer thickness when slips happen, and the recurrence time between slips are analyzed by calculating their complementary cumulative distribution functions (CCDF). Finally, we show the relationship between the friction coefficient drop, kinetic energy release and the drop of gouge layer thickness.

## 4.3. Characteristics of Slips Under Different Normal Loads

The plots presented in Figure 8 and Figure 9 show the general level, changing trend and variations of macroscopic friction coefficient, kinetic energy and gouge layer thickness in both stick and slip phases when subjected to different normal loads. To further compare the frequency and magnitude of slip events, the information of the drop of macroscopic friction coefficient, kinetic energy release, and gouge layer thickness drop when slip occurs, is also extracted by calculating the difference of these quantities between adjacent time step intervals. In the calculations, a threshold of macroscopic friction coefficient drop > 3.0E-5 is used to filter the extremely small events. Then, to avoid including tiny fluctuations of friction coefficient caused by particle rearrangements in the stick phase, a kinetic energy release > 3.0E-8 J is also used to filter the data. These thresholds are chosen after careful analyses of the frequency of all macroscopic friction coefficient drops and kinetic energy releases. Therefore, a slip event is defined when the macroscopic friction coefficient drop and kinetic energy release all satisfy the above thresholds.

To compare the influences on slips caused by different normal loads, the complementary cumulative distribution functions (CCDF) of the slips are calculated. The CCDF is defined as

$$\mathrm{CCDF}(Y) = 1 - \mathrm{CDF}(Y), \tag{15}$$

where $Y$ represents either the friction coefficient drop, kinetic energy release, or the drop of gouge layer thickness, and the CDF is short for cumulative distribution function. The CCDF gives the probability of an event larger than or equal to a certain magnitude, and thus provides a useful tool for comparing the influence of different normal loads on the frequency and magnitude of slip events, especially on the large ones. The CCDFs of macroscopic friction coefficient drop, kinetic energy release, and gouge layer thickness drop are plotted in Figure 10a-c. The different lines in Figure 10 represent the CCDF of slips under different normal loads, where the vertical lines denote the mean magnitudes of the slip events and the red arrow indicates increasing normal load. The farther right side of the CCDF represents a greater number of large magnitude friction coefficient drops, kinetic energy releases, or gouge layer thickness drops have happened under the corresponding normal load.

Figure 10a shows that as the normal load increases, the magnitude of macroscopic friction coefficient drop tends to be smaller. This is consistent with the earlier analyses that higher normal loads constrain the gouge layer more and allow for less change of the contact between particles and plates, which leads to a smaller magnitude of friction coefficient drop. However, the kinetic energy release and gouge layer thickness drop exhibit the opposite trend. Figure 10b demonstrates that as the normal load increases, the associated kinetic energy release increases. This, again, agrees with the seismic moment results (Figure 7) as well as the fact that for higher normal loads, more external work has been done on the fault system and, therefore, more work is converted into kinetic energy during a slip event. The larger magnitude of kinetic energy release induces larger compaction in the gouge layer, as is demonstrated in Figure 10c.

In addition, we can observe in Figure 8 that slips are relatively less frequent with increasing normal load. In order to investigate the influence of normal load on the frequency of slip events, the recurrence time between slips is calculated. The recurrence time is defined as the time between large adjacent slip events, and sequential, small slips are considered as a single large slip. The CCDFs of recurrence time and their means corresponding to different normal loads are presented in Figure 10d, which shows an increasing recurrence time with higher normal loads. This is, again, in response to the phenomena shown in Figure 10b and c in that more external work is done on the fault gouge due to higher normal load. A longer recurrence time, or equivalently longer stick phase results, whereby more elastic energy is stored in the system during the stick phase and more will be released in the form of kinetic energy.

### 4.4. Correlations Between the Stick-Slip Indicators

The friction coefficient, kinetic energy and gouge layer thickness are all useful indicators to interpret the dynamics of granular fault gouge. To better understand the correlation between them, we plot in Figure 11a-e the relationship between the drop of friction coefficient, kinetic energy release and gouge thickness drop for different normal load scenarios. Although scattered, the data generally show linear relationships in log space between these indicators. In other words, under each normal load, as the magnitude of friction coefficient drop increases, there is an increase in the magnitude of both kinetic energy release and gouge layer thickness drop, and the magnitude of gouge layer thickness drop also increases with increasing kinetic energy release, as noted previously.

The correlation coefficients between each pair of parameters are shown in Figure 12 for comparison. The results reveal that the correlation coefficients have an overall increasing trend with increasing normal load. We speculate this is still because for high normal loads the gouge layer is more compact and it behaves like a more unified system when a slip happens, and thus the drop of friction coefficient, kinetic energy release and gouge thickness drop vibrate in a more synchronous manner. Therefore, the kinetic energy release and the drop of friction coefficient and gouge layer thickness occur more simultaneously and yield larger correlation coefficients between them. Among these correlations, the friction coefficient drop and kinetic energy release have the highest correlation, while the friction coefficient drop and gouge thickness drop display the lowest correlation. Because it is widely recognized that it is the frictional weakening that causes the slip event, these correlation coefficients demonstrate that although gouge thickness drop can be observed during slip events, it is not as sensitive as the friction coefficient and kinetic energy. Note that local minima take place at 28 kPa normal load for all three sets of friction coefficients; however, the reason of this is not clear yet. Furthermore, the results shown in Figure 11 is show scatter, and we suspect this could be due to a number of factors, such as gouge thickness, particle size distribution, or potentially, particle-particle and particle-plate friction coefficients. Further investigation will be conducted in the future to give a detailed exploration of these.

### 5. Conclusions and Discussions

FDEM has proven itself to be a useful tool to study the stick-slip behavior of the sheared granular fault gouge. In the FDEM model, the plates and particles are represented by discrete elements to track their motion and interaction with neighboring objects, and each discrete element is further discretized into finite elements to capture its deformation and stress evolution during the shearing stage. The influence of normal load on the dynamics of the sheared granular fault gouge in terms of macroscopic friction coefficient, kinetic energy, gouge layer thickness, and recurrence time between slips has been investigated, and a comparison between the simulated results and laboratory experiments in terms of seismic moment has been given.

The simulation results show that the granular fault system experiences observable deformations in both plates and particles during the shearing phase, and the stresses are concentrated and released periodically in the fault gouge before and after the slip event. The seismic moments from the simulation are qualitatively similar to the laboratory experiments conducted by Geller et al. (2015). The probability density distribution of all seismic moments is consistent with the Gutenberg-Richter law. The NC events are generally distributed with an $M^{-3/2}$ power law scaling and the C events are broadly peaked, concentrated at large $M$ with a log-normal distribution. As the normal load increases, the probability of NC events decreases, while the larger C events become more probable with broader distributions. The kinetic energy of the granular fault system increases with the increasing normal load, while the gouge layer thickness shows the opposite trend and the macroscopic friction coefficient does not experience much change. As the normal load increases, the kinetic energy and gouge layer thickness are less variable. These behaviors may be attributed to the more external work being done on the granular fault gouge when subjected to higher normal load, which results in a more compact gouge layer with smaller variations in gouge thickness and a more energetic system with higher kinetic energy level.

The influence of normal load on the macroscopic friction coefficient drop, kinetic energy release, drop of gouge layer thickness, and the recurrence time between slips, is analyzed by plotting their CCDFs. The results show that as the normal load increases, more slip events with larger kinetic energy release and longer recurrence time occur, and the magnitude of gouge layer thickness drop also tends to be larger. The higher kinetic energy release and larger magnitude of gouge layer thickness drop may be caused by the more external work having been done on the fault system due to the higher normal load and longer recurrence time. As a result, more elastic energy is converted into kinetic energy when a slip happens, and the sudden release of kinetic energy with large magnitude further compacts the gouge layer and causes a large drop of gouge layer thickness. However, the macroscopic friction coefficient drop tends to be smaller with higher normal loads. This is probably because higher normal loads place constraints on the gouge layer and allow only small contact change between particles and plates to take place, and thus lead to a smaller magnitude of friction coefficient change. In addition, the friction coefficient drop, kinetic energy release and the gouge thickness drop for each normal load show good correlations and their correlation coefficients increase with increasing normal load. This tells that the granular fault system exhibits unified behavior when subjected to higher normal loads. Nevertheless, the comparison between these correlation coefficients implies that the gouge thickness is less sensitive than the friction coefficient and kinetic energy in terms of being a stick-slip indicator.

The simulation results are generally in agreement with the experimental and numerical observations obtained earlier (Dorostkar et al., 2017a, 2017b; Ferdowsi et al., 2013; Ferdowsi et al., 2014; Geller et al., 2015). For example, both the two-dimensional laboratory experiment conducted by Geller et al. (2015) and the numerical simulated results here have an approximately linear increase of friction before approaching the peak (Figure 8). However, a nonlinear increase of friction is observed in three-dimensional simulations and experiments (e.g. Dorostkar et al., 2017a; Dorostkar et al., 2017b; Ferdowsi et al., 2014; Rivière et al., 2018; Scuderi et al., 2017b). We suspect this is because the friction in the third dimension provides extra support and leads to a relative steady frictional state before slip event (Hazzard & Mair, 2003). Furthermore, both Geller et al. (2015) experiments and the numerical simulation here show an increasing recurrence time with respect to increasing normal loads, while the biaxial shear laboratory tests such as the one shown in Rivière et al. (2018) demonstrate the opposite trend. The reason may lie in the different driving plate stiffness employed in these investigations: the biaxial shear laboratory tests use stiff steel plates, while here the driving plates are significantly soft ($E = 2.5$ MPa). The detailed influence of plate stiffness on the dynamics of stick-slip events in a granular fault system, together with the above-mentioned unsolved problems, are still open to the geophysics field and further investigations using both laboratory experiments and numerical simulations are necessary to unveil these mysteries. Nevertheless, our results have qualitative representation in the behavior of individual earthquake faults and is consistent with our observations of increased likelihood of large events at higher normal load (Scholz, 2002).

Although the current simulation setup is similar to the laboratory experiments (Geller et al., 2015), because of the different shear velocity used in these two series of investigations, a very detailed quantitative comparison is beyond the scope of current paper. In addition, the kinetic energy reported in this paper is a combination of both plates and gouge since when slip happens, both of them will slip and the kinetic energy of them reveals the dynamic behavior of the entire earthquake system. Because energy plays a pivotal role in the stick-slip performance of granular fault systems, benefiting from the capability of FDEM for detailed stress, strain and motion acquisition, further exploration is being undertaken in order to give a thorough analysis of the kinetic and elastic energy conversion and evolution in particles and plates when subjected to different normal loads. The simulated data will also be evaluated for machine learning purposes to predict the laboratory scale earthquakes, such as the ones that have recently been done (Hulbert et al., 2018; Rouet-Leduc et al., 2018; Rouet-Leduc et al., 2017).

## Acknowledgements

The Los Alamos National Laboratory LDRD [Institutional Support] Program supported this work. Technical support and computational resources from the Los Alamos National Laboratory Institutional Computing Program are highly appreciated. Our data are available by contacting the corresponding author.

## References

Abe, S., & Mair, K. (2005). Grain fracture in 3D numerical simulations of granular shear. *Geophysical Research Letters, 32*(5). doi:10.1029/2004GL022123

Anastasaki, E., Latham, J.-P., & Xiang, J. (2015). Numerical modelling of armour layers with reference to Core-Loc units and their placement acceptance criteria. *Ocean Engineering, 104*(Supplement C), 204-218. doi:10.1016/j.oceaneng.2015.05.010

Annunziata, M. A., Baldassarri, A., Dalton, F., Petri, A., & Pontuale, G. (2016). Increasing 'ease of sliding' also increases friction: when is a lubricant effective? *Journal Of Physics: Condensed Matter, 28*(13), 134001.

Anthony, J. L., & Marone, C. (2005). Influence of particle characteristics on granular friction. *Journal of Geophysical Research: Solid Earth, 110*(B8), B08409. doi:10.1029/2004JB003399

de Arcangelis, L., Ciamarra, M. P., Lippiello, E., & Godano, C. (2011). Micromechanics and statistics of slipping events in a granular seismic fault model. *Journal of Physics: Conference Series, 319*(1), 012001.

Dorostkar, O., Guyer, R. A., Johnson, P. A., Marone, C., & Carmeliet, J. (2017a). On the micromechanics of slip events in sheared, fluid-saturated fault gouge. *Geophysical Research Letters, 44*(12), 6101-6108. doi:10.1002/2017GL073768

Dorostkar, O., Guyer, R. A., Johnson, P. A., Marone, C., & Carmeliet, J. (2017b). On the role of fluids in stick-slip dynamics of saturated granular fault gouge using a coupled computational fluid dynamics-discrete element approach. *Journal of Geophysical Research: Solid Earth, 122*(5), 3689-3700. doi:10.1002/2017JB014099

Dratt, M., & Katterfeld, A. (2017). Coupling of FEM and DEM simulations to consider dynamic deformations under particle load. *Granular Matter, 19*(3), 49. doi:10.1007/s10035-017-0728-3

Ferdowsi, B. (2014). *Discrete element modeling of triggered slip in faults with granular gouge. Application to dynamic earthquake triggering.* (Ph.D Thesis), ETH-Zürich, Switzerland.

Ferdowsi, B., Griffa, M., Guyer, R. A., Johnson, P. A., Marone, C., & Carmeliet, J. (2013). Microslips as precursors of large slip events in the stick-slip dynamics of sheared granular layers: A discrete element model analysis. *Geophysical Research Letters, 40*(16), 4194-4198. doi:10.1002/grl.50813

Ferdowsi, B., Griffa, M., Guyer, R. A., Johnson, P. A., Marone, C., & Carmeliet, J. (2014). Three-dimensional discrete element modeling of triggered slip in sheared granular media. *Physical Review E, 89*(4), 042204.

Gao, K. (2017). *Contributions to Tensor based Stress Variability Characterisation in Rock Mechanics.* (Ph.D thesis), University of Toronto, Canada.

Gao, K., & Lei, Q. (2018). Influence of boundary constraints on stress heterogeneity modelling. *Computers And Geotechnics, 99*, 130-136. doi:10.1016/j.compgeo.2018.03.003

Geller, D. A., Ecke, R. E., Dahmen, K. A., & Backhaus, S. (2015). Stick-slip behavior in a continuum-granular experiment. *Physical Review E, 92*(6), 060201.

Griffa, M., Daub, E. G., Guyer, R. A., Johnson, P. A., Marone, C., & Carmeliet, J. (2011). Vibration-induced slip in sheared granular layers and the micromechanics of dynamic earthquake triggering. *EPL (Europhysics Letters), 96*(1), 14001.

Griffa, M., Ferdowsi, B., Daub, E. G., Guyer, R. A., Johnson, P. A., Marone, C., & Carmeliet, J. (2012). Meso-mechanical analysis of deformation characteristics for dynamically triggered slip in a granular medium. *Philosophical Magazine, 92*(28-30), 3520-3539. doi:10.1080/14786435.2012.700417

Griffa, M., Ferdowsi, B., Guyer, R. A., Daub, E. G., Johnson, P. A., Marone, C., & Carmeliet, J. (2013). Influence of vibration amplitude on dynamic triggering of slip in sheared granular layers. *Physical Review E, 87*(1), 012205.

Guises, R., Xiang, J., Latham, J.-P., & Munjiza, A. (2009). Granular packing: numerical simulation and the characterisation of the effect of particle shape. *Granular Matter, 11*(5), 281-292. doi:10.1007/s10035-009-0148-0

Guo, L., Latham, J.-P., & Xiang, J. (2015). Numerical simulation of breakages of concrete armour units using a three-dimensional fracture model in the context of the combined finite-discrete element method. *Computers & Structures, 146*(Supplement C), 117-142. doi:10.1016/j.compstruc.2014.09.001

Gutenberg, B., & Richter, C. F. (1955). Magnitude and Energy of Earthquakes. *Nature, 176*, 795. doi:10.1038/176795a0

Hazzard, J. F., & Mair, K. (2003). The importance of the third dimension in granular shear. *Geophysical Research Letters, 30*(13). doi:10.1029/2003GL017534

Hulbert, C., Rouet-Leduc, B., Ren, C. X., Riviere, J., Bolton, D. C., Marone, C., & Johnson, P. A. (2018). Estimating the Physical State of a Laboratory Slow Slipping Fault from Seismic Signals. *arXiv preprint arXiv:1801.07806*.

Johnson, P. A., Ferdowsi, B., Kaproth, B. M., Scuderi, M., Griffa, M., Carmeliet, J., . . . Marone, C. (2013). Acoustic emission and microslip precursors to stick-slip failure in sheared granular material. *Geophysical Research Letters, 40*(21), 5627-5631. doi:10.1002/2013GL057848

Johnson, P. A., & Jia, X. (2005). Nonlinear dynamics, granular media and dynamic earthquake triggering. *Nature, 437*(7060), 871-874.

Johnson, P. A., Savage, H., Knuth, M., Gomberg, J., & Marone, C. (2008). Effects of acoustic waves on stick-slip in granular media and implications for earthquakes. *Nature, 451*(7174), 57-60.

Knight, E. E., Rougier, E., & Lei, Z. (2015). *Hybrid optimization software suite (HOSS) - educational version. Technical Report* (LA-UR-15-27013). Los Alamos National Laboratory

Latham, J.-P., Anastasaki, E., & Xiang, J. (2013a). New modelling and analysis methods for concrete armour unit systems using FEMDEM. *Coastal Engineering, 77*(Supplement C), 151-166. doi:10.1016/j.coastaleng.2013.03.001

Latham, J.-P., Mindel, J., Xiang, J., Guises, R., Garcia, X., Pain, C., . . . Munjiza, A. (2009). Coupled FEMDEM/Fluids for coastal engineers with special reference to armour stability and breakage. *Geomechanics and Geoengineering, 4*(1), 39-53. doi:10.1080/17486020902767362

Latham, J.-P., & Munjiza, A. (2012). Porosity and packing simulations of particles with any shape or size – dynamic 3D results *Coastal Engineering 2002* (pp. 1424-1435): World Scientific Publishing Company.

Latham, J.-P., Munjiza, A., Garcia, X., Xiang, J., & Guises, R. (2008a). Three-dimensional particle shape acquisition and use of shape library for DEM and FEM/DEM simulation. *Minerals Engineering, 21*(11), 797-805. doi:10.1016/j.mineng.2008.05.015

Latham, J.-P., Munjiza, A., Mindel, J., Xiang, J., Guises, R., Garcia, X., . . . Piggott, M. (2008b). Modelling of massive particulates for breakwater engineering using coupled FEMDEM and CFD. *Particuology, 6*(6), 572-583. doi:10.1016/j.partic.2008.07.010

Latham, J.-P., Xiang, J., Belayneh, M., Nick, H. M., Tsang, C.-F., & Blunt, M. J. (2013b). Modelling stress-dependent permeability in fractured rock including effects of propagating and bending fractures. *International Journal of Rock Mechanics and Mining Sciences, 57*, 100-112.

Latham, J. P., & Munjiza, A. (2004). The Modelling of Particle Systems with Real Shapes. *Philosophical Transactions: Mathematical, Physical and Engineering Sciences, 362*(1822), 1953-1972.

Lei, Q., & Gao, K. (2018). Correlation between Fracture Network Properties and Stress Variability in Geological Media. *Geophysical Research Letters*. doi:10.1002/2018GL077548

Lei, Q., Latham, J.-P., Tsang, C.-F., Xiang, J., & Lang, P. (2015). A new approach to upscaling fracture network models while preserving geostatistical and geomechanical characteristics. *Journal of Geophysical Research, 120*(7), 4784-4807. doi:10.1002/2014JB011736

Lei, Q., Latham, J.-P., Xiang, J., & Tsang, C.-F. (2017). Role of natural fractures in damage evolution around tunnel excavation in fractured rocks. *Engineering Geology, 231*, 100-113. doi:10.1016/j.enggeo.2017.10.013

Lei, Z., Rougier, E., Knight, E. E., & Munjiza, A. (2013, 2013/1/1/). *Block Caving Induced Instability Analysis using FDEM.* Paper presented at the 47th U.S. Rock Mechanics/Geomechanics Symposium, San Francisco, USA.

Lei, Z., Rougier, E., Knight, E. E., & Munjiza, A. (2014). A framework for grand scale parallelization of the combined finite discrete element method in 2d. *Computational Particle Mechanics, 1*(3), 307-319. doi:10.1007/s40571-014-0026-3

Lei, Z., Rougier, E., Knight, E. E., Munjiza, A. A., & Viswanathan, H. (2016). A generalized anisotropic deformation formulation for geomaterials. *Computational Particle Mechanics, 3*(2), 215-228. doi:10.1007/s40571-015-0079-y

Lisjak, A., Grasselli, G., & Vietor, T. (2014). Continuum–discontinuum analysis of failure mechanisms around unsupported circular excavations in anisotropic clay shales. *International Journal of Rock Mechanics and Mining Sciences, 65*(Supplement C), 96-115. doi:10.1016/j.ijrmms.2013.10.006

Lisjak, A., Liu, Q., Zhao, Q., Mahabadi, O. K., & Grasselli, G. (2013). Numerical simulation of acoustic emission in brittle rocks by two-dimensional finite-discrete element analysis. *Geophysical Journal International, 195*(1), 423-443. doi:10.1093/gji/ggt221

Ma, G., Zhou, W., Chang, X.-L., & Chen, M.-X. (2016). A hybrid approach for modeling of breakable granular materials using combined finite-discrete element method. *Granular Matter, 18*(1), 7. doi:10.1007/s10035-016-0615-3

Mair, K., & Abe, S. (2008). 3D numerical simulations of fault gouge evolution during shear: Grain size reduction and strain localization. *Earth And Planetary Science Letters, 274*(1), 72-81. doi:10.1016/j.epsl.2008.07.010

Marone, C. (1998). The effect of loading rate on static friction and the rate of fault healing during the earthquake cycle. *Nature, 391*(6662), 69-72.

Marone, C., Raleigh, C. B., & Scholz, C. H. (1990). Frictional behavior and constitutive modeling of simulated fault gouge. *Journal of Geophysical Research: Solid Earth, 95*(B5), 7007-7025. doi:10.1029/JB095iB05p07007

MiDi, G. (2004). On dense granular flows. *The European Physical Journal E, 14*(4), 341-365.

Munjiza, A. (1992). *Discrete elements in transient dynamics of fractured media.* (PhD Thesis), Swansea University.

Munjiza, A., & Andrews, K. (1998). NBS contact detection algorithm for bodies of similar size. *International Journal For Numerical Methods In Engineering, 43*(1), 131-149.

Munjiza, A., Rougier, E., & John, N. W. M. (2006). MR linear contact detection algorithm. *International Journal For Numerical Methods In Engineering, 66*(1), 46-71. doi:10.1002/nme.1538

Munjiza, A. A. (2004). *The Combined Finite-Discrete Element Method*: John Wiley & Sons.

Munjiza, A. A., Knight, E. E., & Rougier, E. (2011). *Computational Mechanics of Discontinua*: John Wiley & Sons.

Munjiza, A. A., Lei, Z., Divic, V., & Peros, B. (2013a). Fracture and fragmentation of thin shells using the combined finite–discrete element method. *International Journal For Numerical Methods In Engineering, 95*(6), 478-498. doi:10.1002/nme.4511

Munjiza, A. A., Rougier, E., & Knight, E. E. (2014). *Large strain finite element method: a practical course*: John Wiley & Sons.

Munjiza, A. A., Rougier, E., Knight, E. E., & Lei, Z. (2013b). *HOSS: An integrated platform for discontinua simulations.* Paper presented at the Frontiers of Discontinuous Numerical Methods and Practical Simulations in Engineering and Disaster Prevention.

Passelègue, F. X., Schubnel, A., Nielsen, S., Bhat, H. S., Deldicque, D., & Madariaga, R. (2016a). Dynamic rupture processes inferred from laboratory microearthquakes. *Journal of Geophysical Research: Solid Earth, 121*(6), 4343-4365. doi:10.1002/2015JB012694

Passelègue, F. X., Spagnuolo, E., Violay, M., Nielsen, S., Di Toro, G., & Schubnel, A. (2016b). Frictional evolution, acoustic emissions activity, and off-fault damage in simulated faults sheared at seismic slip rates. *Journal of Geophysical Research: Solid Earth, 121*(10), 7490-7513. doi:10.1002/2016JB012988

Pica Ciamarra, M., Lippiello, E., de Arcangelis, L., & Godano, C. (2011). Statistics of slipping event sizes in granular seismic fault models. *EPL, 95*(5), 54002.

Reccia, E., Cazzani, A., & Cecchi, A. (2012). FEM-DEM modeling for out-of-plane loaded masonry panels: a limit analysis approach. *Open Civil Engineering Journal, 6*(1), 231-238.

Rivière, J., Lv, Z., Johnson, P. A., & Marone, C. (2018). Evolution of b-value during the seismic cycle: Insights from laboratory experiments on simulated faults. *Earth And Planetary Science Letters, 482*, 407-413. doi:10.1016/j.epsl.2017.11.036

Rouet-Leduc, B., Hulbert, C., Bolton, D. C., Ren, C. X., Riviere, J., Marone, C., . . . Johnson, P. A. (2018). Estimating Fault Friction from Seismic Signals in the Laboratory. *Geophysical Research Letters*. doi:10.1002/2017GL076708

Rouet-Leduc, B., Hulbert, C., Lubbers, N., Barros, K., Humphreys, C. J., & Johnson, P. A. (2017). Machine Learning Predicts Laboratory Earthquakes. *Geophysical Research Letters, 44*(18), 9276-9282. doi:10.1002/2017GL074677

Rougier, E., & Munjiza, A. A. (2010). *MRCK_3D contact detection algorithm.* Paper presented at the Proceedings of 5th international conference on discrete element methods, London, UK.

Scholz, C. H. (2002). *The mechanics of earthquakes and faulting* (2$^{nd}$ ed.): Cambridge University Press.

Scuderi, M. M., Collettini, C., & Marone, C. (2017a). Frictional stability and earthquake triggering during fluid pressure stimulation of an experimental fault. *Earth And Planetary Science Letters, 477*(Supplement C), 84-96. doi:10.1016/j.epsl.2017.08.009

Scuderi, M. M., Collettini, C., Viti, C., Tinti, E., & Marone, C. (2017b). Evolution of shear fabric in granular fault gouge from stable sliding to stick slip and implications for fault slip mode. *Geology, 45*(8), 731-734. doi:10.1130/G39033.1

Tatone, B. S. A., & Grasselli, G. (2015). A calibration procedure for two-dimensional laboratory-scale hybrid finite–discrete element simulations. *International Journal of Rock Mechanics and Mining Sciences, 75*(Supplement C), 56-72. doi:10.1016/j.ijrmms.2015.01.011

Trivino, L. F., & Mohanty, B. (2015). Assessment of crack initiation and propagation in rock from explosion-induced stress waves and gas expansion by cross-hole seismometry and FEM–DEM method. *International Journal of Rock Mechanics and Mining Sciences, 77*(Supplement C), 287-299. doi:10.1016/j.ijrmms.2015.03.036

Viswanathan, H. S., Carey, J. W., Karra, S., Porter, M. L., Rougier, E., Currier, R. P., . . . Hyman, J. D. (2015). *Integrated Experimental and Computational Study of Hydraulic Fracturing and the Use of Alternative Fracking Fluids.* Paper presented at the 49th U.S. Rock Mechanics/Geomechanics Symposium, San Francisco, USA.

Vyazmensky, A., Stead, D., Elmo, D., & Moss, A. (2010). Numerical Analysis of Block Caving-Induced Instability in Large Open Pit Slopes: A Finite Element/Discrete Element Approach. *Rock Mechanics and Rock Engineering, 43*(1), 21-39. doi:10.1007/s00603-009-0035-3

Wang, C., Elsworth, D., & Fang, Y. (2017). Influence of weakening minerals on ensemble strength and slip stability of faults. *Journal of Geophysical Research: Solid Earth, 122*(9), 7090-7110. doi:doi:10.1002/2016JB013687

Xu, D., Kaliviotis, E., Munjiza, A., Avital, E., Ji, C., & Williams, J. (2013). Large scale simulation of red blood cell aggregation in shear flows. *Journal Of Biomechanics, 46*(11), 1810-1817. doi:10.1016/j.jbiomech.2013.05.010

**Table Captions**

Table 1. Material and numerical simulation parameters.

**Figure Captions**

Figure 1. Numerical representation of the fault containing granular fault gouge using the DEM and FDEM. (a) In the DEM the plates are simplified by a set of bonded particles and the gouge layer is composed of a series of rigid particles. (b) In the FDEM the plate is explicitly represented and both plates and particles are further discretized into finite elements to capture their deformation and stress evolution.

Figure 2. Schematic of contact interaction and contact force calculation.

Figure 3. FDEM model of the granular fault system and its geometrical dimensions. The top right inset shows the discretized particle used in the simulation that composed of 24 approximately equal size elements and its comparison with a circle.

Figure 4. Deformation of the granular fault system subjected to 44 kPa normal load: (a) final state of the model, i.e. at the end of shear; (b) overlay of the outlines of the initial (i.e. at the beginning of shear) and final states of the model according to their coordinates.

Figure 5. Evolution of major principal stress $\sigma_1$ (compression positive) in the granular fault gouge immediate (a) before and (b) after a slip event for the fault subjected to 44 kPa normal load. (Note that only part of the fault gouge is presented and locations of these two stages on the time series plot is further marked in Figure 8e).

Figure 6. Average slip $M_j/N_j$ with respect to $M_j$ for different normal loads. Note that different marker shapes represent different normal loads; the color markers represent NC events and the black markers denote C events. The events for $P = 20\text{-}44$ kPa are offset vertically for better visualization.

Figure 7. Probability density distribution of the seismic moment of (a) all events, (b) NC events and (c) C events for different normal loads.

Figure 8. The time series of macroscopic friction coefficient, kinetic energy, and gouge layer thickness of the granular fault system sheared under a constant shear velocity of 0.5 mm/s and subjected to various normal loads (a) $P = 12$ kPa, (b) $P = 20$ kPa, (c) $P = 28$ kPa, (d) $P = 36$ kPa, and (e) $P = 44$ kPa. Note that positions of the circle and star in (e) correspond to Figure 5a & b, respectively, and the inset of (e) is a zoom in of the friction coefficient change immediately before and after the slip event shown in Figure 5.

Figure 9. Boxplots of the (a) macroscopic friction coefficient, (b) kinetic energy and (c) gouge layer thickness for granular fault gouge subjected to various normal loads.

Figure 10. The complementary cumulative distribution function (CCDF) of (a) macroscopic friction coefficient drop, (b) kinetic energy release, (c) gouge layer thickness drop, and (d) recurrence time between slips for the granular fault gouge subjected to different normal loads. The vertical lines denote the mean magnitudes corresponding to each normal load and the red arrow indicates increasing normal load.

Figure 11. The correlations between macroscopic friction coefficient drop, kinetic energy release, and gouge layer thickness drop for different normal loads: (a) $P = 12$ kPa, (b) $P = 20$ kPa, (c) $P = 28$ kPa, (d) $P = 36$ kPa, and (e) $P = 44$ kPa. (Note that KE denotes kinetic energy).

Figure 12. The correlation coefficients between the logarithms of the macroscopic friction coefficient drop, kinetic energy release and gouge layer thickness drop.

Table 1. Material and numerical simulation parameters.

| Property | Value | Property | Value |
| --- | --- | --- | --- |
| Particle diameter | 1.2 or 1.6 mm | Stiff bar density | 2,800 kg/m$^3$ |
| Particle density | 1,150 kg/m$^3$ | Stiff bar Young's modulus | 30 GPa |
| Particle Young's modulus | 0.4 GPa | Stiff bar Poisson's ratio | 0.33 |
| Particle Poisson's ratio | 0.4 | Foam density | 1,150 kg/m$^3$ |
| Particle-particle friction coefficient | 0.15 | Foam Young's modulus | 1 MPa |
| Number of particles | 2,817 | Foam Poisson's ratio | 0.4 |
| Main plate density | 1,150 kg/m$^3$ | Contact penalty | 4 GPa |
| Main plate Young's modulus | 2.5 MPa | Time step | 1.0E-7 s |
| Main plate Poisson's ratio | 0.49 | Normal load $P$ | 12-44 kPa |
| Particle-plate friction coefficient | 0.15 | Shear velocity $V$ | 0.5 mm/s |

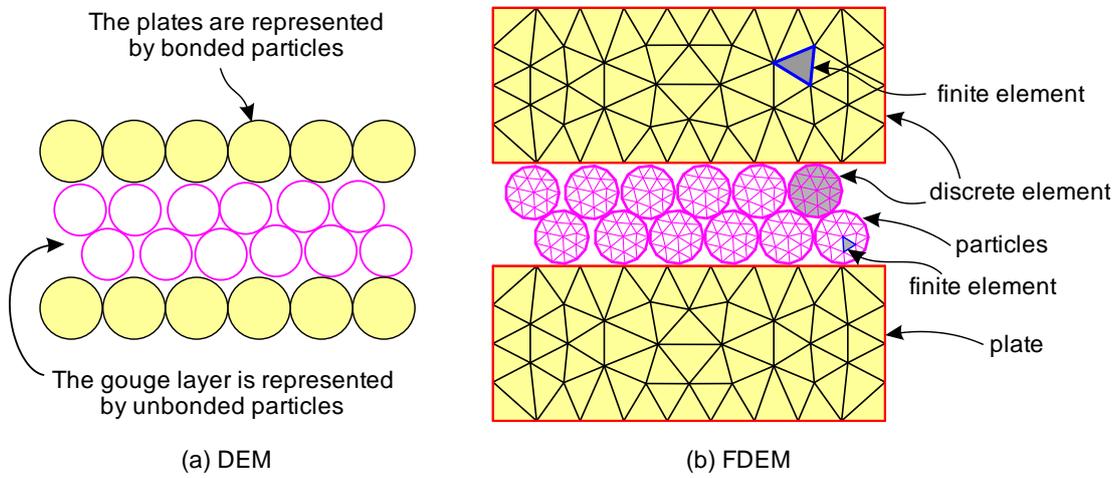

Figure 1. Numerical representation of the fault containing granular fault gouge using the DEM and FDEM. (a) In the DEM the plates are simplified by a set of bonded particles and the gouge layer is composed of a series of rigid particles. (b) In the FDEM the plate is explicitly represented and both plates and particles are further discretized into finite elements to capture their deformation and stress evolution.

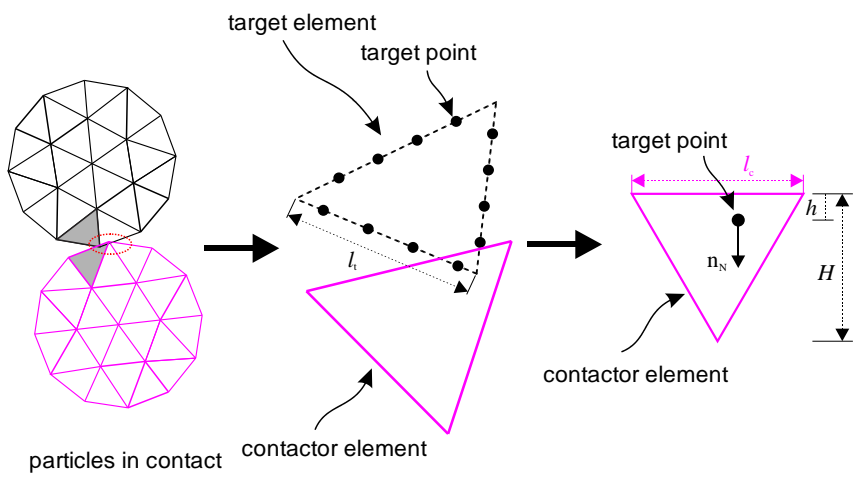

Figure 2. Schematic of contact interaction and contact force calculation.

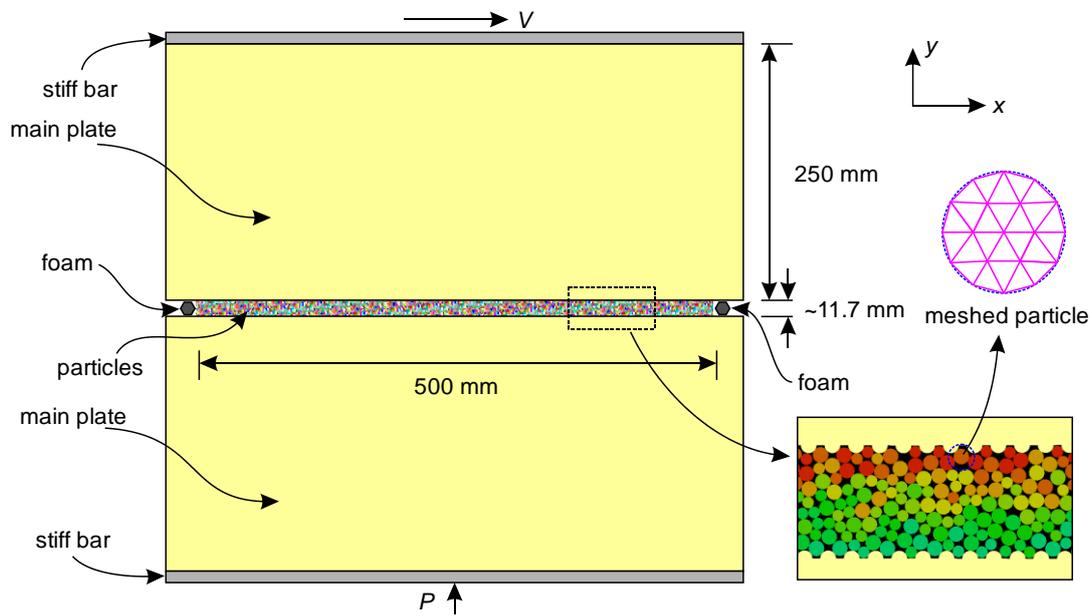

Figure 3. FDEM model of the granular fault system and its geometrical dimensions. The top right inset shows the discretized particle used in the simulation that composed of 24 approximately equal size elements and its comparison with a circle.

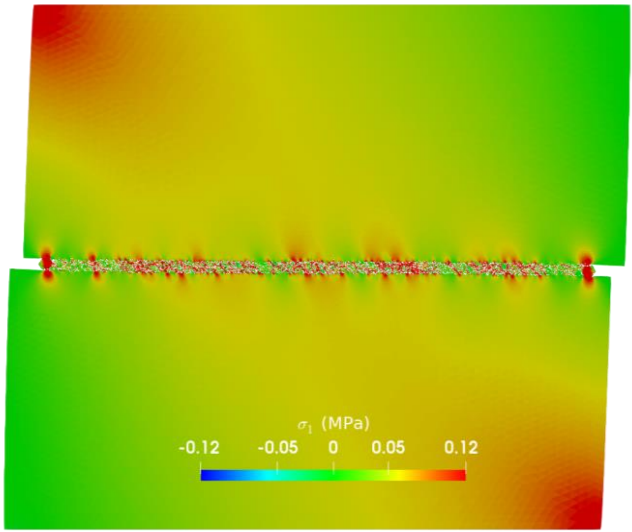

(a) Deformation and stress distribution ($\sigma_1$, compression positive) at the final state (end of shear)

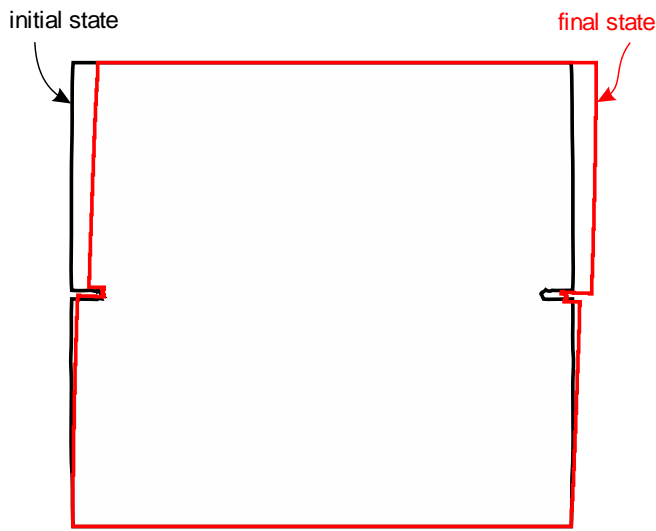

(b) Overlay of the outlines of beginning and final state

Figure 4. Deformation of the granular fault system subjected to 44 kPa normal load: (a) final state of the model, i.e. at the end of shear; (b) overlay of the outlines of the initial (i.e. at the beginning of shear) and final states of the model according to their coordinates.

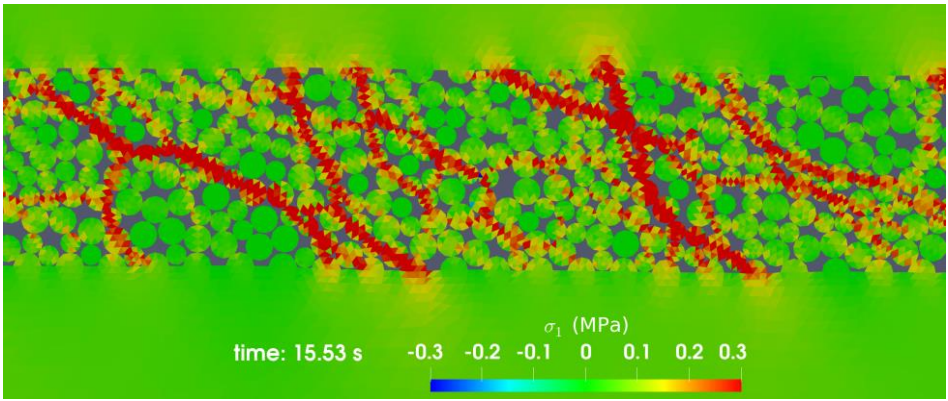

(a) Stress concentration in the granular fault gouge immediately before a slip event

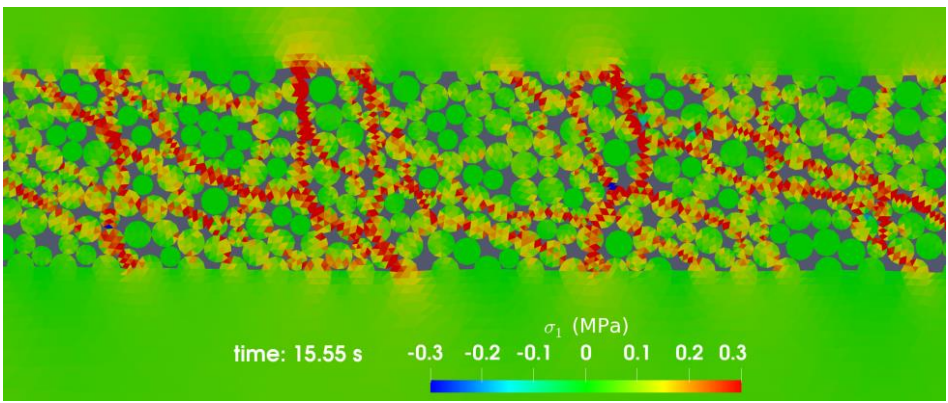

(b) Relatively uniform stress distribution in the granular fault gouge immediately after the slip event

Figure 5. Evolution of major principal stress $\sigma_1$ (compression positive) in the granular fault gouge immediate (a) before and (b) after a slip event for the fault subjected to 44 kPa normal load. (Note that only part of the fault gouge is presented and locations of these two stages on the time series plot is further marked in Figure 8e).

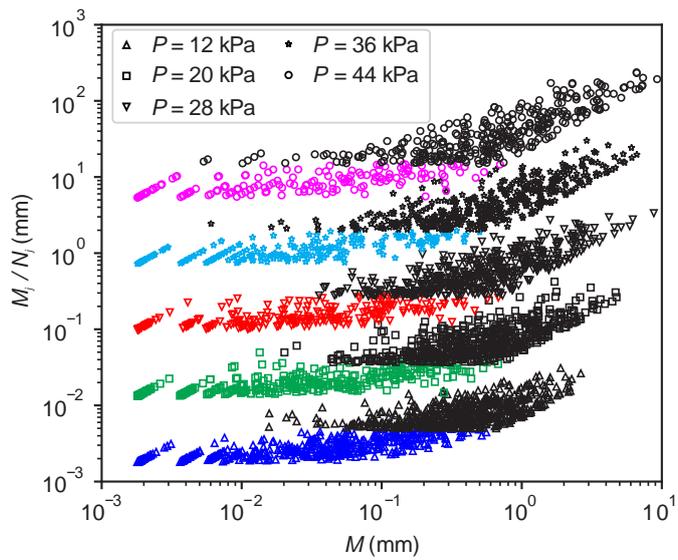

Figure 6. Average slip $M_j/N_j$ with respect to $M_j$ for different normal loads. Note that different marker shapes represent different normal loads; the color markers represent NC events and the black markers denote C events. The events for $P$ = 20-44 kPa are offset vertically for better visualization.

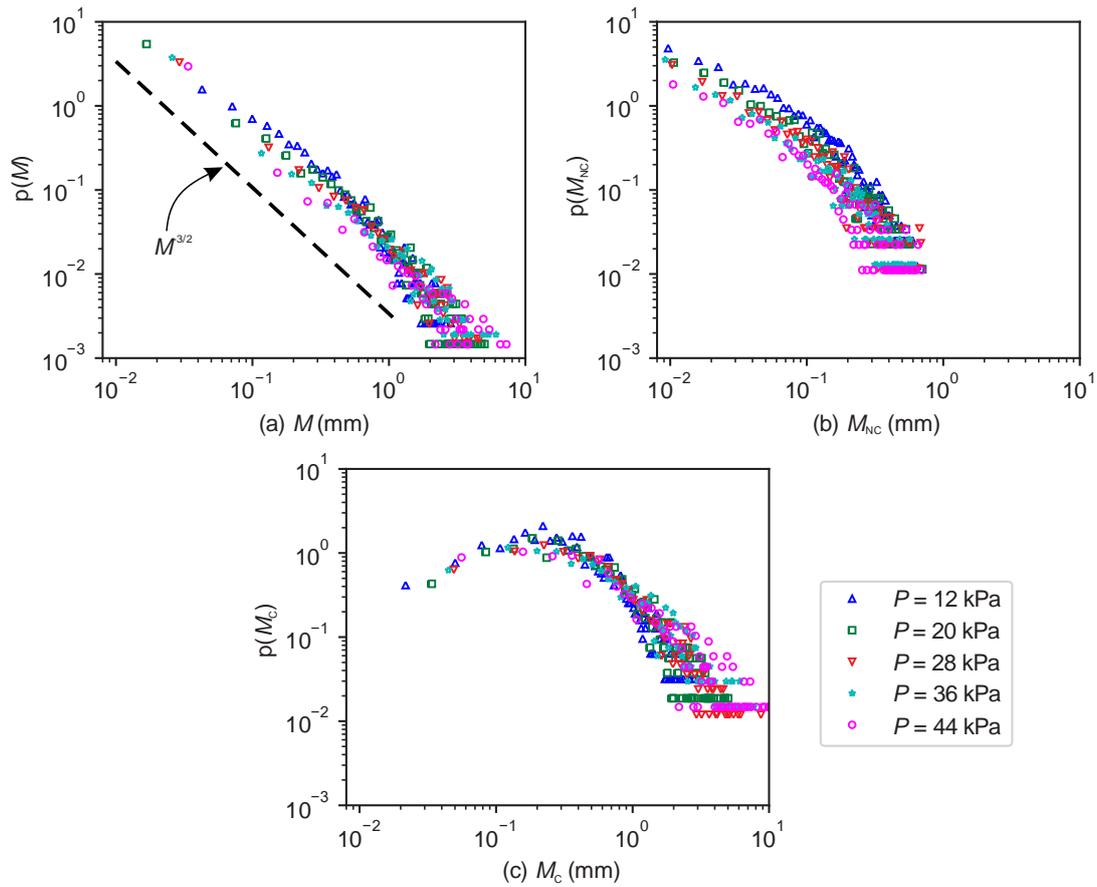

Figure 7. Probability density distribution of the seismic moment of (a) all events, (b) NC events and (c) C events for different normal loads.

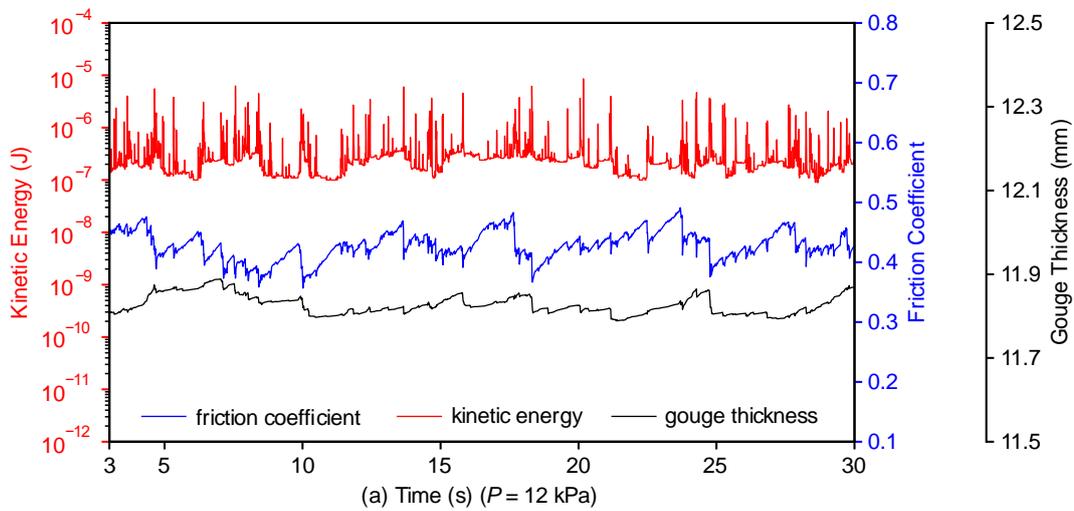
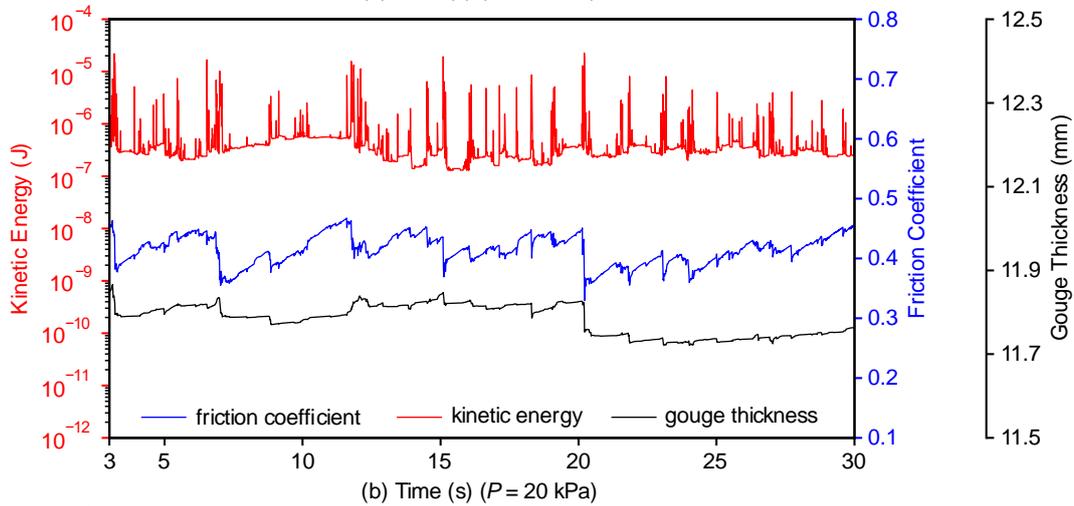
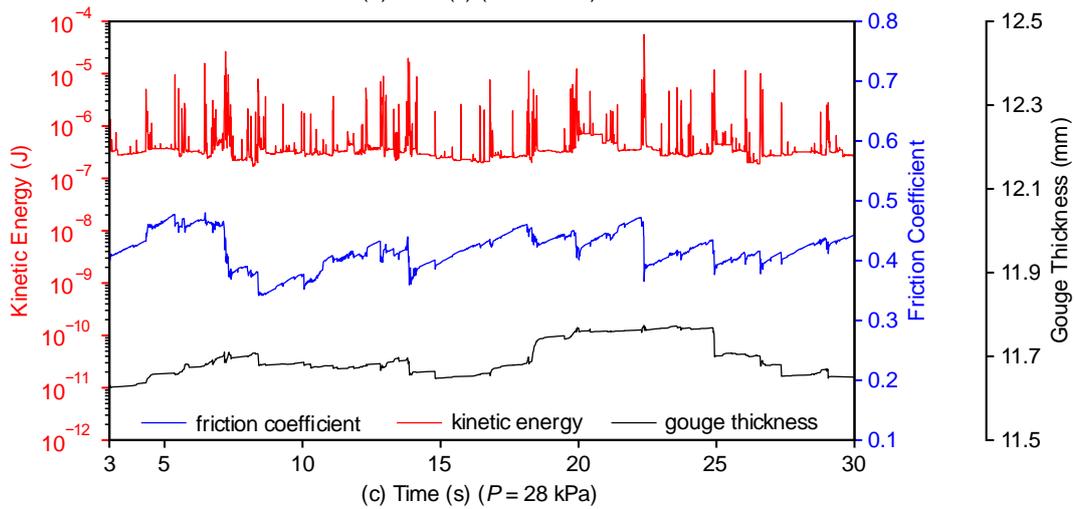

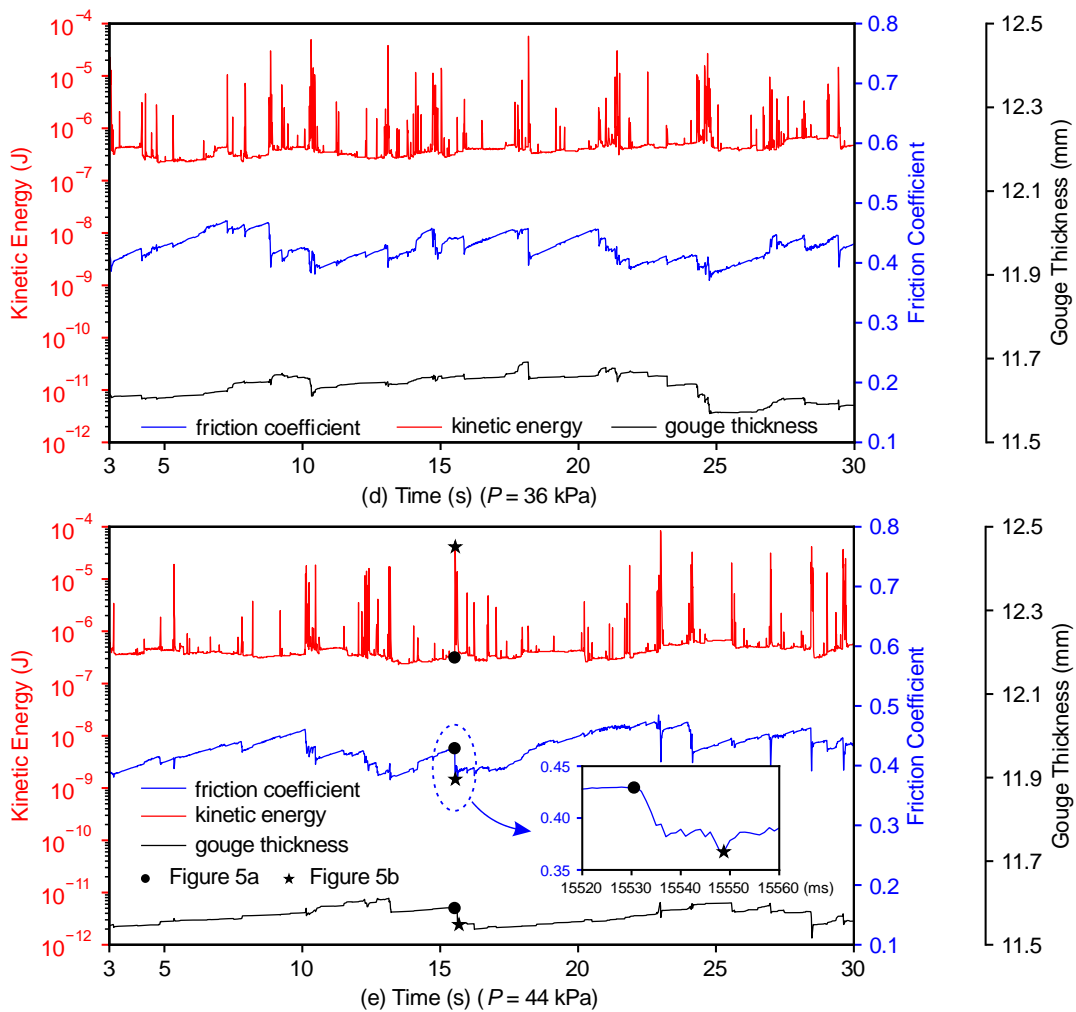

Figure 8. The time series of macroscopic friction coefficient, kinetic energy, and gouge layer thickness of the granular fault system sheared under a constant shear velocity of 0.5 mm/s and subjected to various normal loads (a) $P$ = 12 kPa, (b) $P$ = 20 kPa, (c) $P$ = 28 kPa, (d) $P$ = 36 kPa, and (e) $P$ = 44 kPa. Note that positions of the circle and star in (e) correspond to Figure 5a & b, respectively, and the inset of (e) is a zoom in of the friction coefficient change immediately before and after the slip event shown in Figure 5.

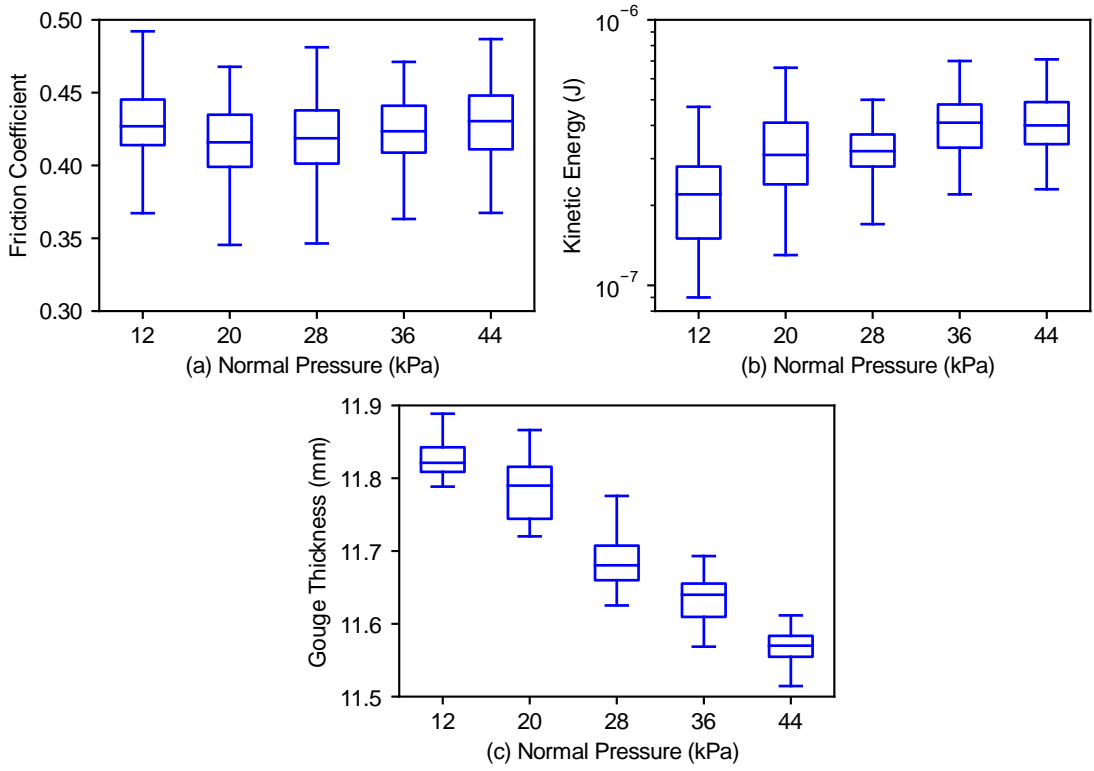

Figure 9. Boxplots of the (a) macroscopic friction coefficient, (b) kinetic energy and (c) gouge layer thickness for granular fault gouge subjected to various normal loads.

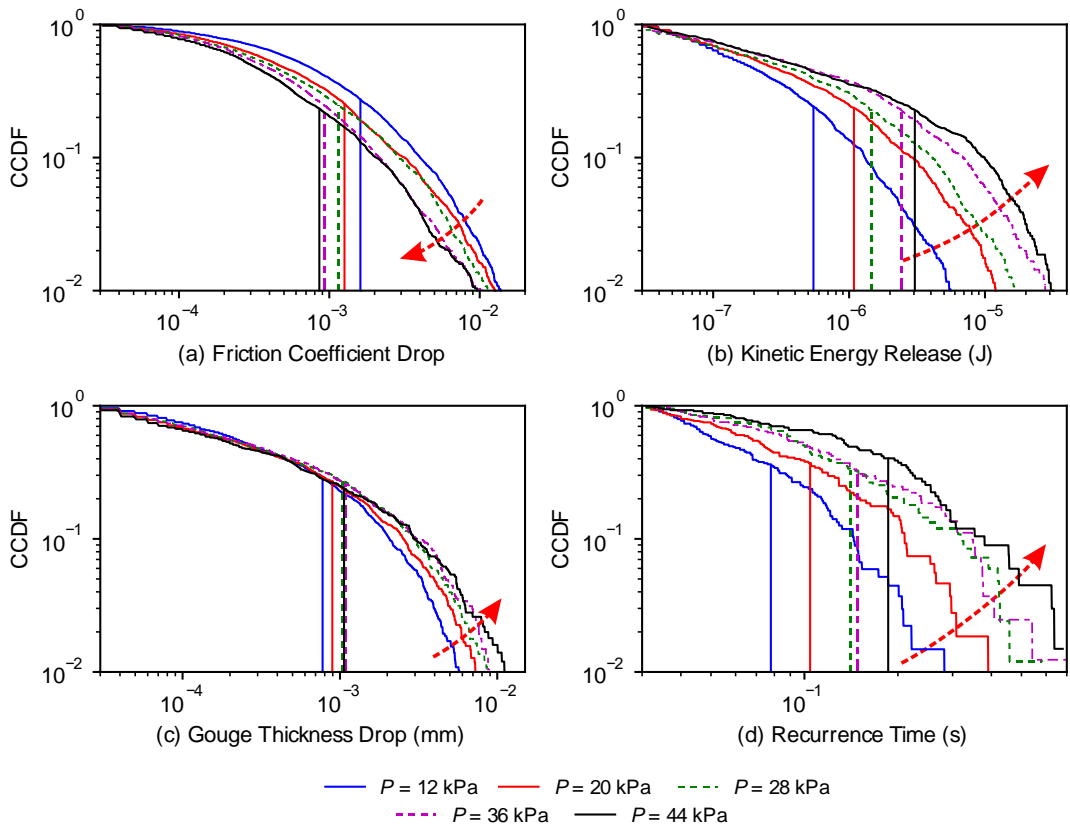

Figure 10. The complementary cumulative distribution function (CCDF) of (a) macroscopic friction coefficient drop, (b) kinetic energy release, (c) gouge layer thickness drop, and (d) recurrence time between slips for the granular fault gouge subjected to different normal loads. The vertical lines denote the mean magnitudes corresponding to each normal load and the red arrow indicates increasing normal load.

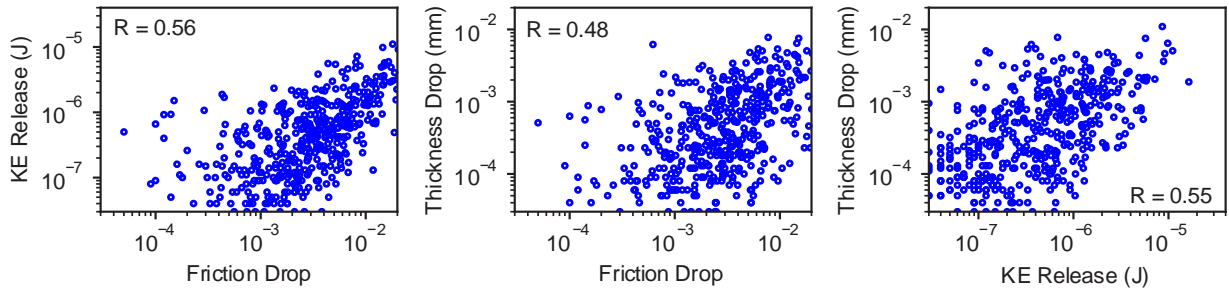
(a) $P$ = 12 kPa

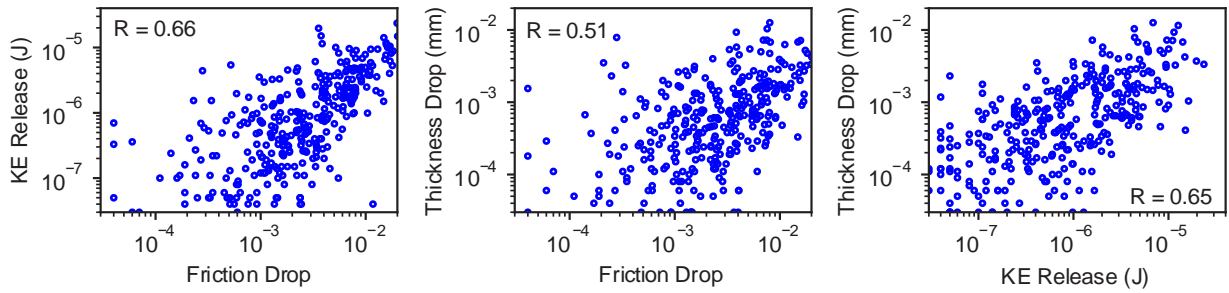
(b) $P$ = 20 kPa

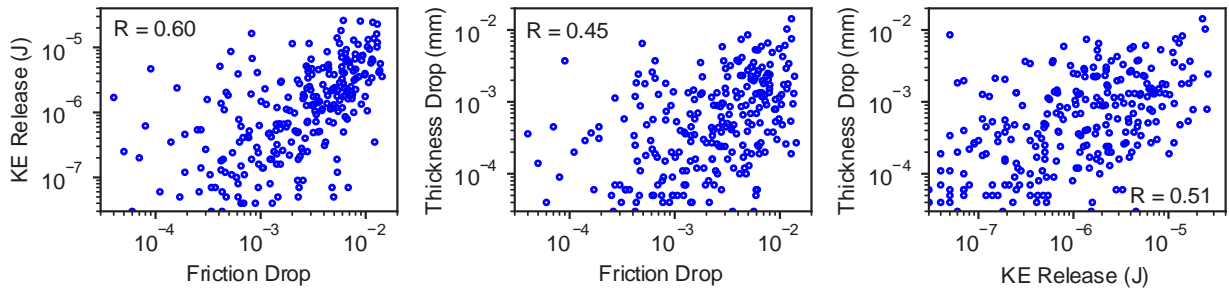
(c) $P$ = 28 kPa

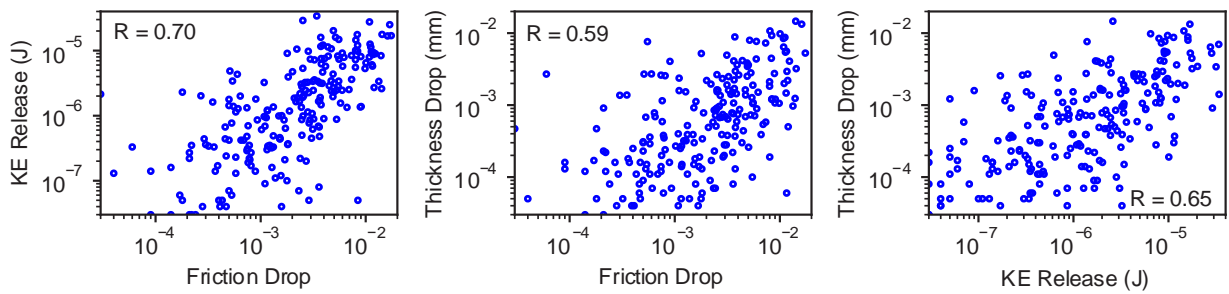
(d) $P$ = 36 kPa

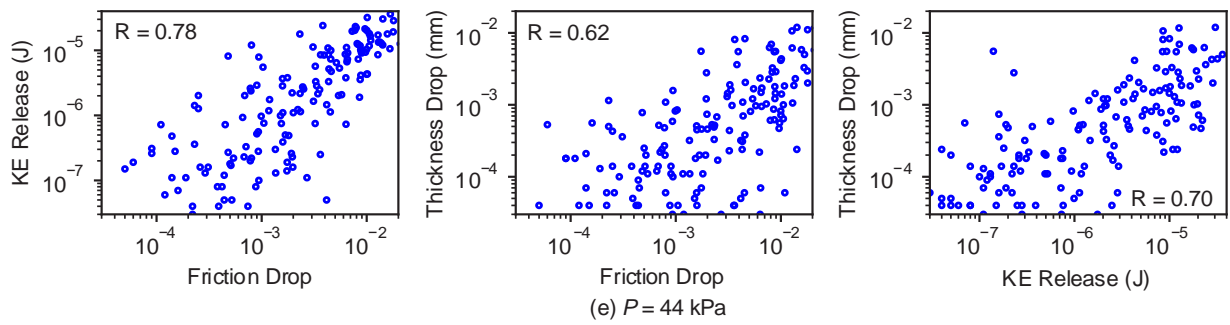

(e) *P* = 44 kPa

Figure 11. The correlations between macroscopic friction coefficient drop, kinetic energy release, and gouge layer thickness drop for different normal loads: (a) *P* = 12 kPa, (b) *P* = 20 kPa, (c) *P* = 28 kPa, (d) *P* = 36 kPa, and (e) *P* = 44 kPa. (Note that KE denotes kinetic energy).

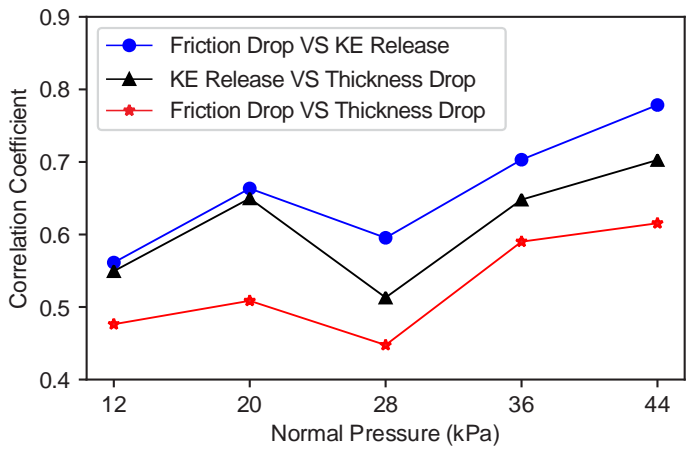

Figure 12. The correlation coefficients between the logarithms of the macroscopic friction coefficient drop, kinetic energy release and gouge layer thickness drop.